\documentclass[epjST]{svjour}
\usepackage{graphics}
\usepackage{graphics,graphicx,dcolumn,bm,fleqn,epic,eepic,float}
\usepackage{amssymb,amsmath,multirow,rotate,rotating,color}
\usepackage[sort&compress,square,comma,numbers]{natbib}
\usepackage{color}
\usepackage{upgreek}
\usepackage[utf8]{inputenc}

\newcommand{\sound}{c_{\text s}}

\newcommand{\figref}[1]{Fig.~\ref{fig:#1}}
\newcommand{\eqnref}[1]{Eq.~(\ref{eq:#1})}

\newcommand{\revisedtext}[1]{#1}

\begin{document}
\title{Recent advances in the simulation of particle-laden flows}
\author{Jens Harting\inst{1,2}\fnmsep\thanks{\email{j.harting@tue.nl}}\and Stefan Frijters\inst{1} \and Marco Ramaioli\inst{3}\and Martin Robinson\inst{4}\and Dietrich E. Wolf\inst{5} \and Stefan Luding\inst{6}}
\institute{
Department of Applied Physics, Eindhoven University of Technology, P.O. Box 513,\\ 5600MB Eindhoven, The Netherlands \and 
Faculty of Science and Technology, Mesa+ Institute, University of
Twente, P.O. Box 217,\\ 7500AE Enschede, The Netherlands
%\and Nestl\'e Research Center, Route du Jorat 57, 1000 Lausanne 26, Switzerland
\and Environmental Hydraulics Laboratory, École Polytechnique Fédérale de Lausanne, 1015 Lausanne, Switzerland
\and Mathematical Institute, University of Oxford, Andrew Wiles Building,\\ Radcliffe Observatory Quarter,
Woodstock Road, Oxford OX2 6GG, United Kingdom
\and  Department of Physics, University of Duisburg-Essen, Lotharstr.1, 47057 Duisburg,\\ Germany
\and MSM, MESA+, CTW, Department of Engineering Technology, University of Twente,\\ P.O. Box 217, 7500AE Enschede, The Netherlands}
\abstract{
A substantial number of algorithms exists for the simulation of moving particles
suspended in fluids. However, finding 
the best method to address a particular physical problem
is often highly non-trivial and depends on the properties of the particles and
the involved fluid(s) together. In this report we provide a short overview on a
number of existing simulation methods and provide two state of the art
examples in more detail. 
In both cases, the particles are described using a Discrete Element
Method (DEM). The DEM solver is usually coupled to a fluid-solver, which can
be classified as grid-based or mesh-free (one example for each is given). Fluid
solvers feature different resolutions relative to the particle size and
separation. First, a multicomponent lattice Boltzmann algorithm (mesh-based and
with rather fine resolution) is presented to study the behavior of particle stabilized
fluid interfaces and second, a Smoothed Particle Hydrodynamics implementation
(mesh-free, meso-scale resolution, similar to the particle size) 
is introduced to highlight a new player in the field, 
which is expected to be particularly suited for flows including free
surfaces.
} %end of abstract
\maketitle

\section{Introduction}
Particles suspended in fluids are common in our daily life. Examples include
human blood (which is to good approximation made of particulate red blood cells
suspended in plasma) \cite{dupin07,Aidun2010,kruger2011efficient,bib:jens-janoschek-toschi-2010b,goldsmith75}, 
or tooth paste and wall paint, which are mixtures of finely
ground solid ingredients in fluids, as well as e.g.\ geophysical debris 
flow \cite{leonardi14b}. Extreme examples are the sand on a beach
or in a desert, when blown away by the wind \cite{sauermann00,araujo13}, or industrial gas phase synthesis of nanoparticles~\cite{lorke2012}. Also every-day food-powders as we use them in our kitchen, 
related to the dispersion of instant-drinks \cite{robinson2013dispersion}, 
are challenging examples of particle-fluid systems.

Long-range fluid-mediated hydrodynamic interactions often dictate the behavior
of particle-fluid mixtures. The majority of analytical results for the particle
scale behavior of suspensions has been obtained in the regime of vanishing
Reynolds numbers (viscous flow) since an analytical treatment of the full
Navier-Stokes equations is generally difficult or even impossible. 
Computer simulation methods are indispensable for many-particle systems,
for the inclusion of inertia effects (Reynolds numbers $> 1$) and Brownian
motion (Peclet number of order $1$). These systems often contain various 
important time scales that can differ by many orders of magnitude,
but nevertheless have to be resolved by the simulation, 
leading to multiscale problems that can only be solved by a large numerical effort. 

Many simulation methods have been developed to simulate particle-fluid
mixtures, including the early works by pioneers as Tsuji and Herrmann
\cite{tsuji85,tanaka93,kalthoff96,sauermann00}. 
Examples for more recent reviews on research on multi-phase and particle fluid
flow models, including their experimental validation and approaches
at different scales of resolution are provided in Refs.~
\cite{chu08numerical,link05flow,deen07review,vanderHoef08numerical,robinson2013ijmf,araujo13,srivastava14,leonardi14,leonardi14b,ge14,guo14}
and references therein.
All of them have their inherent strengths but also some disadvantages
and can generally be divided into two separate classes: The methods in the
first category restrict the treatment of the fluid to approximations of the
hydrodynamic interactions between suspended particles only, while the methods
in the second category try to tackle all degrees of freedom of the involved fluids by approximating the Navier-Stokes equations.
%SL Jens: did you want to say of the particles in the fluid? or of the fluid between 
% the particles???  

The simplest approach is Brownian Dynamics which does not contain
long-ranged hydrodynamic interactions among particles at
all~\cite{Huetter00}. Brownian Dynamics with full hydrodynamic
interactions utilizes a mobility matrix which is based on tensor
approximations which are exact in the limit of zero Reynolds number and
zero particle volume fraction~\cite{Petera99,Ahlrichs01}. However,
the computational effort scales with the cube of the particle number due
to the inversion of matrices.  Pair-Drag simulations have been proposed by
Silbert et al.~\cite{Silbert97a}, and include hydrodynamic interactions in an
approximative way. They have focused on suspensions with high densities (up to
$50\,\%$) of uncharged spherical colloidal particles.  Stokesian Dynamics has
been presented by Bossis and Brady in the 80s and applied subsequently e.g. in~\cite{bossis84a,sierou01a,phung96a,brady:88}. 
%For example, Melrose and
%Ball have performed detailed studies of shear thickening colloids using
%Stokesian Dynamics simulations~\cite{melrose04e}. 
However, as the previous
examples, this method is limited to Reynolds numbers close to zero. The
inclusion of long-range hydrodynamic interactions causes the computational
effort to become very high for dynamical simulations with many particles and
makes an efficient parallelization difficult. The method is still widely used
due to its physical motivation and its robustness.  Boundary-element methods
are more flexible than Stokesian dynamics and can also be used to simulate
non-spherical or deformable particles, but the computational effort is even
higher~\cite{loewenberg96}.

As stated above, these methods represent the hydrodynamic interactions between
suspended particles in an approximative way. If the level of approximation is
not well chosen, it might lead to unphysical artifacts, as was shown for
example by Knudsen et al.~\cite{knudsen-werth-wolf:2008}, who studied a system
where electrostatic repulsion and hydrodynamic interactions between particles were both
important. They found that even for dilute systems some existing models
overestimate the effect of hydrodynamic damping leading effectively to an attraction
of like-charged particles. 

Furthermore, all the methods listed above assume that hydrodynamic interactions
are fully developed and that the dynamics of the fluid and of the solved
particles can be treated separately. In reality, this is not the case.
Hydrodynamic interactions are time dependent due to local stresses at the
fluid-particle interfaces. A number of more recent methods attempt to describe
the time dependent long-range hydrodynamics properly with the computational
effort scaling linearly with the number of particles. These include recent
mesoscopic methods like Dissipative Particle Dynamics \cite{espanol-warren},
the lattice Boltzmann method
\cite{bib:ladd-verberg:2001,bib:jens-komnik-herrmann:2004}, Stochastic Rotation
Dynamics/Multi Particle Collision
Dynamics~\cite{gompper-mpc-review:2008,A7:Malev99,
A7:Malev00,bib:jens-hecht-ihle-herrmann:2005,bib:jens-hecht-bier-reinshagen-herrmann:2006},
or Smoothed Particle
Hydrodynamics~\cite{robinson2013ijmf,robinson2013sph,robinson2013dispersion}.
However, for small Reynolds numbers, the computational gain of these methods is
lost due to the additional effort needed to describe the motion of the fluid.

Finite element or finite difference methods need a proper meshing of the
computational domain which is not trivial for complicated boundary conditions
as in the case of dense suspensions. Therefore, many authors only simulated a
limited number of static configurations rather than the full dynamics of the
system.  Advances in remeshing techniques as well as more powerful computers
have allowed to overcome these problems. 
Also, in order to avoid remeshing at all, uniform grids can often be
used~\cite{Fogelson88,Hoefler99,Hoefler00-a,Fonseca04a,Fonseca04b}.
Another recent class of simulation methods involves moving grids
see Refs.\ \cite{srivastava14,guo14,ge14} and references therein.

If the particles are very massive and the density of the fluid is very
low, a full hydrodynamic treatment of the solvent is not needed anymore.
Then, a coarse-grained description of the fluid where the fluid is resolved on a length scale larger than the particles might be sufficient. 
Much larger systems can be treated this way, but the
coarse-graining is justified only in certain situations. An example is
the pneumatic transport of a powder in a pipe which is a common
process in many industrial applications~\cite{tanaka93,MCN0003,Strauss1,Strauss2}.

In the remainder of this paper we give examples, where simplified approaches to
treat hydrodynamic interactions between suspended particles are supposed to
fail. First, we review recent work on the interplay of particle laden flows
with fluid interfaces. The simulation method of choice here is based on a
Discrete Element Method (DEM) and a multiphase lattice Boltzmann algorithm,
where the fluid is resolved considerably better than the particles.
Then, we introduce an application of a relatively new player in the field, i.e.
a combined DEM and Smoothed Particle Hydrodynamics approach, which has its
strength in the simulation of systems containing free fluid interfaces, while leaving
the fluid "meso-resolved", on a scale slightly larger than the particle diameter.

\section{Particles in multicomponent fluid flows}

\subsection{The lattice Boltzmann -- Discrete Element Method}

The use of particles as an alternative to surfactants for the stabilization of
emulsions is very attractive, such as in the food, cosmetics, and
medical industries, where they are used to stabilize barbecue sauces and sun cremes,
or to produce sophisticated ways to deliver drugs at the right position in the
human body.  The microscopic processes leading to the commercial interest can
be understood by assuming an oil-water mixture. Without any additives, the
phases would seperate and the oil would float on top of the water. Adding small
particles, however, causes these particles to diffuse to the interface, which is
then stabilized due to a reduced surface energy. If, for example, individual
droplets of one phase are covered by particles, such systems are also referred
to as ``Pickering emulsions''~\cite{bib:ramsden:1903,bib:pickering:1907}.
Particularly interesting properties of such emulsions are the blocking of
Ostwald ripening and their complex rheology due to irreversible particle
adsorption at interfaces.

Computer simulations are a promising route to understanding the dynamics of particle
stabilized emulsions. However, the shortcomings of traditional simulation
methods quickly become obvious: a suitable algorithm is not only required to
deal with simple fluid dynamics but has to be able to simulate several fluid
species while also considering the motion of the particles and the
fluid-particle interactions. Some recent approaches trying to solve these
problems utilize the lattice Boltzmann method (LBM) for the description of the
solvents~\cite{bib:succi:2001}. The LBM is an alternative to conventional
Navier-Stokes solvers and is well established in the literature. It is
attractive for the current application since a number of multiphase and
multicomponent models exist which are comparably straightforward to
implement~\cite{bib:shan-chen:1993,bib:shan-chen:1994,bib:swift-orlandini-osborn-yeomans:1996,bib:lishchuk-care-halliday:2003}.
In addition, the method has been combined with a discrete elements method (DEM)
to simulate arbitrarily shaped particles in flow and is commonly used to study
the behavior of particle-laden single phase
flows~\cite{bib:ladd-verberg:2001,bib:jens-komnik-herrmann:2004,bib:jens-herrmann-bennaim:2008}.

A few groups combined multiphase lattice Boltzmann solvers with the known
algorithms for suspended
particles~\cite{bib:stratford-adhikari-pagonabarraga-desplat-cates:2005,bib:joshi-sun:2009,bib:jens-jansen-2011,bib:jens-frijters-2012a}.
Here, an approach based on the multicomponent lattice Boltzmann model of
Shan and Chen is used which allows the simulation of multiple
fluid components with surface
tension~\cite{bib:shan-chen:1993,bib:jens-jansen-2011,bib:jens-frijters-2012a,bib:jens-floriang-2011}.
The model generally allows arbitrary movements and rotations of arbitrarily
shaped hard shell particles.  Further, it allows an arbitrary choice of the
particle wettability -- one of the most important parameters for the dynamics
of multiphase suspensions~\cite{bib:binks-horozov:2006,bib:binks:2002}.

In the LBM, the fluid is treated as a
cluster of pseudo-particles that move on a lattice under the action of external
forces. 
A distribution function $f_i$ is associated to each pseudo-particle, which denotes the 
probability to find the particle at a position
${\bf r}$ and with a velocity in direction ${\bf
e}_i$. Position and velocity spaces are both discretised,
where $\Delta x$ defines the grid spacing, and ${\bf e}_i$ are the discretised velocity
directions. Here, we use a 3D lattice defined by 19 discrete velocities (D3Q19).
Together with the discrete time step $\Delta t$, the
time evolution of $f_i$ is governed by the so-called lattice Boltzmann
equation,
\begin{equation}
\label{eq:lbe}
f_i({\bf r} + {\bf e}_i \Delta t, t + \Delta t) - f_i({\bf r}, t) = \Omega_i.
\end{equation}
The left-hand side of Eq.~\ref{eq:lbe} denotes advection, while the right hand-side is given by the collision operator
$\Omega_i$ specifying the collision rate between the fluid pseudo-particles. From now on, we set $\Delta x$, $\Delta t$ and the unit mass to unity for convenience. A simple
approximation is the Bhatnagar-Gross-Krook (BGK) operator, $\Omega_i =
-(f_i - f_i^{\text{eq}})/ \tau$, which describes the relaxation of $f_i$
towards its local equilibrium, $f_i^{\text{eq}}$, on a time scale $\tau$. The
relaxation time defines the macroscopic dynamical viscosity $\eta$ via
$\eta = \rho \sound^2\frac{\Delta x^2}{\Delta t}\left(\tau - \frac{1}{2}
\right)$, where $\sound = 1 / \sqrt{3}$ is the lattice speed of sound. The
equilibrium distribution is given by a truncated and discretized Maxwell-Boltzmann
distribution~\cite{bib:succi:2001}.

In the Shan-Chen multi-component model~\cite{bib:shan-chen:1993}, each species $\sigma$ of $k$ miscible or
immiscible fluids is described by a set of
distribution functions $f_{\sigma,i}$. As
a consequence, $k$ lattice Boltzmann equations with relaxation times
$\tau_\sigma$ have to be iterated. The interaction between the fluid
species is mediated via a local force density,
\begin{equation}
\label{eq:shanchen}
{\bf F}_{\sigma}({\bf r},t) = -\Psi_\sigma({\bf r}, t) \sum_{\sigma'} g_{\sigma \sigma'} \sum_{{\bf r}'} \Psi_{\sigma'}({\bf r}',t)({\bf r}' - {\bf r}),
\end{equation}
which is then added to the right hand side of Eq.~\ref{eq:lbe}.
Here, the contributions of all species $\sigma'$ are taken into account, and
the force density acting on species $\sigma$ is obtained. The sum runs over all
neighbouring sites ${\bf r}'$ of site ${\bf r}$.  $g_{\sigma \sigma'}$ is a
coupling constant defining the interaction strength between species $\sigma$
and $\sigma'$. It is related to the surface tension between both species. The
pseudopotential $\Psi_\sigma$ is a function of the density $\rho_\sigma$, in
our case $\Psi_\sigma = 1 - \exp(- \rho_\sigma)$. It defines the equation of
state of the multicomponent system. The Shan-Chen model belongs to the class of
front capturing methods. It is suitable to track interfaces for which the
topology evolves in time, for example, the deformation or even breakup of a
droplet in a shear flow -- as in the current section.

If there is no clear scale separation between immersed particles and the
lateral interface extension, it may be necessary to model the particles
explicitly. Here, we consider an ensemble of particles with well-defined
wetting behaviour. The particles themselves may have a complex shape and can be
supplied with a constitutive model (of which the simplest is a rigid particle
model as we use it here).  There are several approaches to simulate a system of
two immiscible fluids and particles. One example which has recently been
applied by several groups is to couple Molecular Dynamics (MD) or the Discrete Element Method (DEM) to the
LBM~\cite{bib:jens-jansen-2011,bib:jens-frijters-2012a,bib:jens-floriang-2011,bib:stratford-adhikari-pagonabarraga-desplat-cates:2005,bib:kim-stratford-adhikari-cates:2008,bib:joshi-sun:2010,bib:joshi-sun:2009}.
The advantage of this combination is the possibility to resolve the arbitrarily
shaped particles as well as both fluids in such a way that all relevant
hydrodynamical properties are included.

The suspended particles are discretised on the lattice of the fluid solver.
Following the approach originally proposed by Ladd \cite{bib:ladd-verberg:2001}, populations
propagating from fluid to particle sites are bounced back in the direction they
came from. Then they receive additional momentum due to the motion of the
particles. In order to satisfy the local conservation laws, linear and angular
momentum contributions are assigned to the corresponding particles as well.
These in turn are used to update the particle positions and orientations. In
the multi-component algorithm used here, the outermost sites covered by a
particle are filled with a virtual fluid corresponding to a suitable average of
the surrounding unoccupied sites. This approach provides accurate dynamics of
the two-component fluid near the particle surface. The wetting properties of
the particle surface can be controlled by shifting the local density difference
of both fluid species which leads to a well-defined contact angle $\theta_p$ on the
particle surface~\cite{bib:jens-jansen-2011,bib:jens-floriang-2011,bib:jens-frijters-2012a}.

All simulation algorithms for suspensions combining a DEM for the suspended
particles with a discretized solver of the underlying Navier-Stokes dynamics
suffer from the finite resolution of the fluid solver: when two particles
approach each other, there is always a distance below which the fluid solver is
not able to resolve the hydrodynamic interactions anymore. This generally
causes a depletion interaction between particles which one can overcome by
adding a lubrication correction to the DEM part of the algorithm which acts on
very small
distances~\cite{Nguyen2002,bib:jens-jansen-2011,bib:jens-floriang-2011,bib:jens-janoschek-toschi-2014}.

\subsection{Dimensionless numbers}
\label{sec:deformation}
Here, we present as an example an application where the particle-laden flow
needs to be treated by an algorithm which is able to handle moderate Reynolds
numbers as well as multiple fluid phases. This section represents a summary of
some results given in more detail in a recent article by Frijters et
al.~\cite{bib:jens-frijters-2012a} and the reader is referred to this article
for details of the implementation and simulation parameters. A fraction $\chi$
of the interface of a droplet is covered with particles while a shear flow is
imposed on a second fluid surrounding the droplet. This shear flow causes the
droplet to deform -- a process which is common in the transport of particle
stabilized emulsions or their production process.  The deformation of the
droplet can be quantified using the dimensionless deformation parameter
\begin{equation}
  \label{eq:taylor}
  D \equiv \frac{L-B}{L+B}
\end{equation}
introduced by Taylor~\cite{bib:taylor:1932,bib:taylor:1934}, where $L$ is the
length and $B$ is the breadth of the droplet. A deformed droplet has lost
its spherical shape and thus gains a preferred alignment. 
%This is expressed through
%the inclination angle $\theta_d$. 
In the limit of small deformations, Taylor predicts a linear dependence of the
deformation of a droplet on the capillary number (see Eq.~\ref{eq:Ca-eff}),
with a particularly simple form for equiviscous fluids, i.e. $D=35/32
\mathrm{Ca}$. Due to the finite system size, we define the capillary and Reynolds numbers in terms of an effective shear rate $\dot{\gamma}^{\mathrm{eff}}$, which is measured at the interface of the droplet during the simulation instead of the shear rate $\dot{\gamma}$ applied via the moving boundaries:
% ($\lambda \equiv \mu_d / \mu_m = 1$):
%\begin{equation}
%  \label{eq:taylor-deformation}
%  D = \frac{19 \mu_d + 16 \mu_m}{16 \mu_d + 16 \mu_m}  \mathrm{Ca} = \frac{35}{32} \mathrm{Ca}
%\end{equation}
\begin{equation}
  \label{eq:Ca-eff}
  \mathrm{Ca}^{\mathrm{eff}} \equiv \frac{\mu_m \dot{\gamma}^{\mathrm{eff}} R_d}{\sigma}
  \hbox{,}\qquad
%\end{equation}
%\begin{equation}
%  \label{eq:Re-eff}
  \mathrm{Re}^{\mathrm{eff}} \equiv \frac{\rho_m \dot{\gamma}^{\mathrm{eff}} R_d^2}{\mu_m}
  \hbox{,}
\end{equation}
where $\mu_m$ is the dynamic viscosity of the medium, $R_d$ is the radius of
the initial -- undeformed, hence spherical -- droplet and $\sigma$ is the
surface tension. We
also define the ratio of the droplet and medium viscosity $\lambda \equiv \mu_d
/ \mu_m = 1$ in all presented data. The effective Reynolds number
is varied between approximately $0.6 < \mathrm{Re}^{\mathrm{eff}} < 25$.

\subsection{Distribution of nanoparticles}
\label{ssec:distribution}
To understand the effect of nanoparticles on the deformation properties of the droplet, we first discuss how they position themselves at and move over the droplet interface as the droplet is sheared.
\begin{figure}
\begin{center}
\begin{tabular}{l l l l }
a)\vspace*{-0.3cm} & & & \\
&\includegraphics[width=0.18\linewidth]{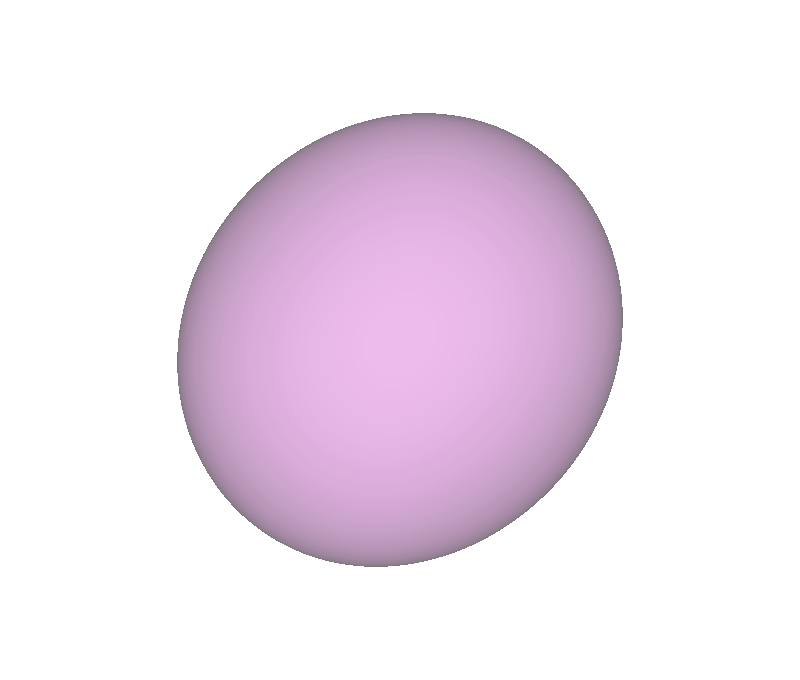} &
\includegraphics[width=0.18\linewidth]{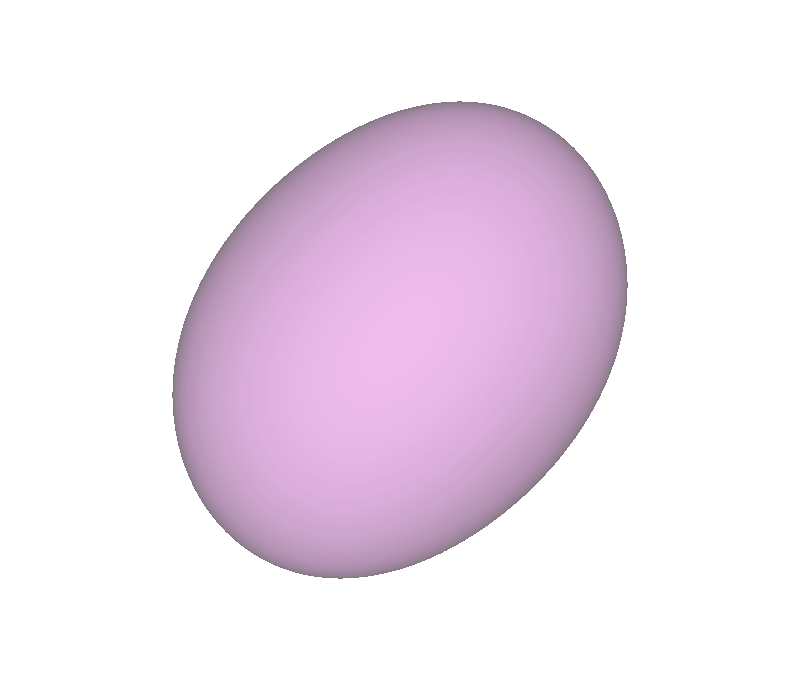} &
\includegraphics[width=0.18\linewidth]{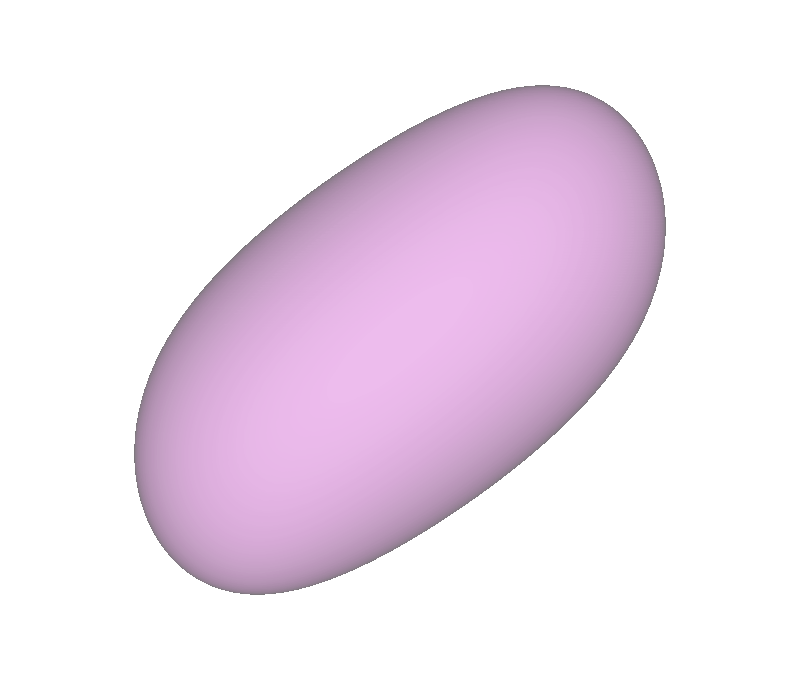} \\
b)\vspace*{-0.3cm} & & & \\
&\includegraphics[width=0.18\linewidth]{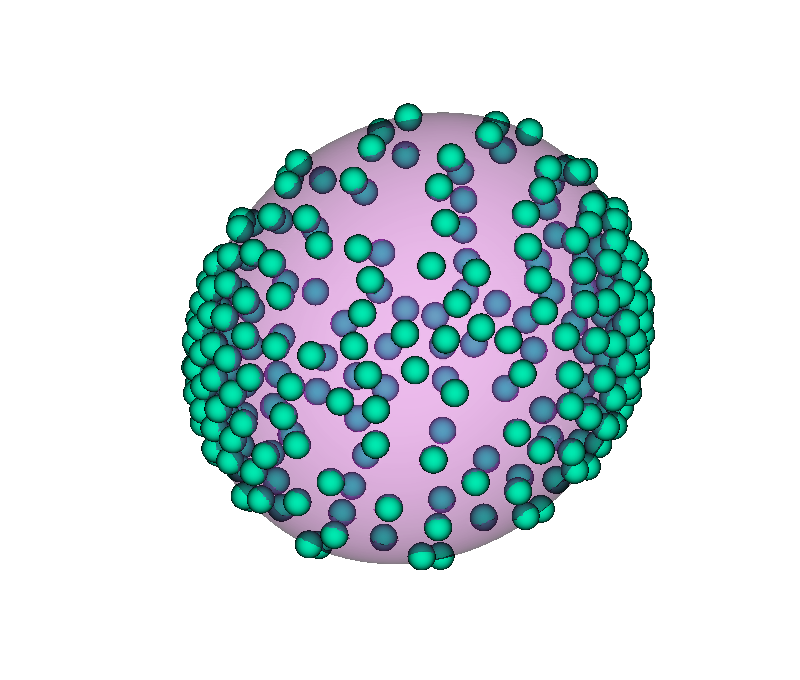} &
\includegraphics[width=0.18\linewidth]{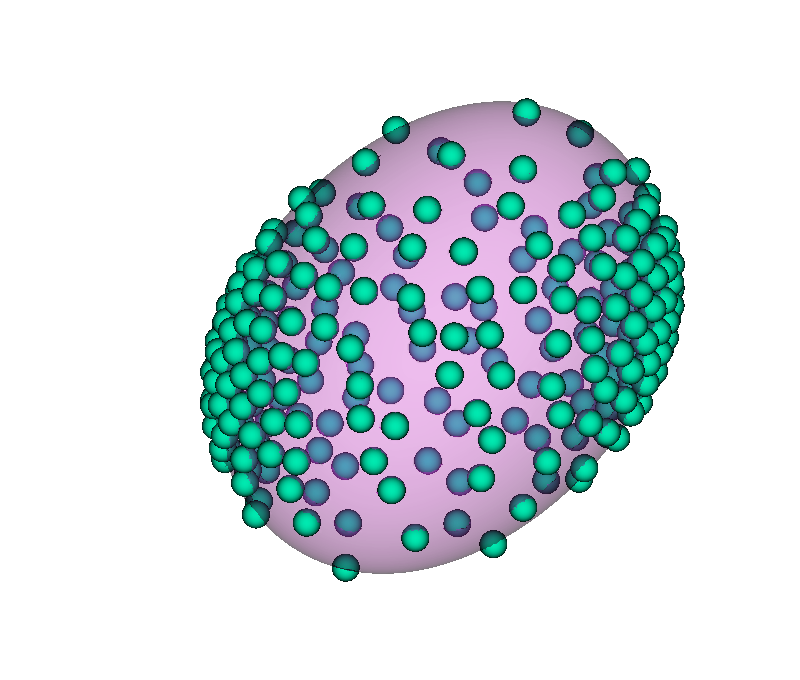} &
\hspace{-0.4cm}\includegraphics[width=0.18\linewidth]{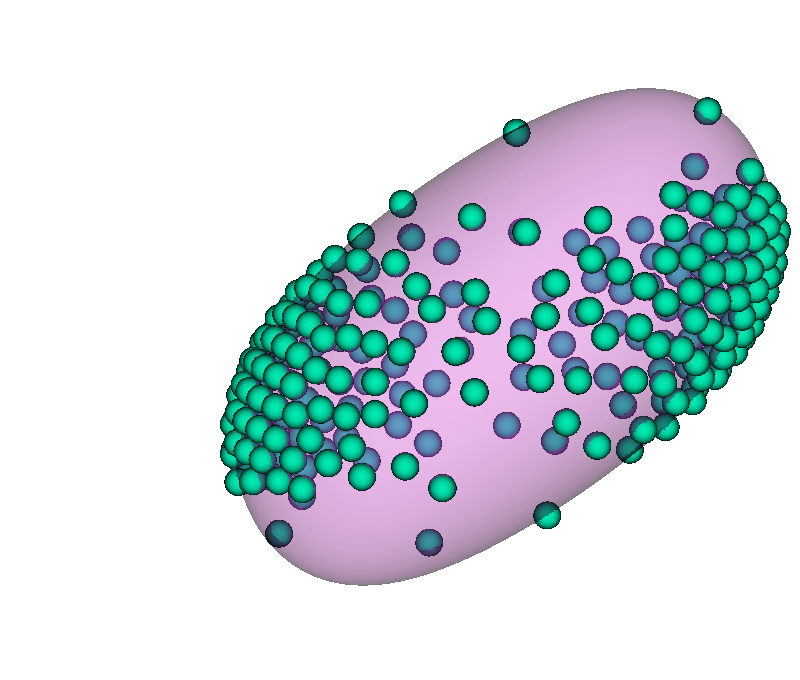} \\
c)\vspace*{-0.3cm} & & & \\
&\includegraphics[width=0.18\linewidth]{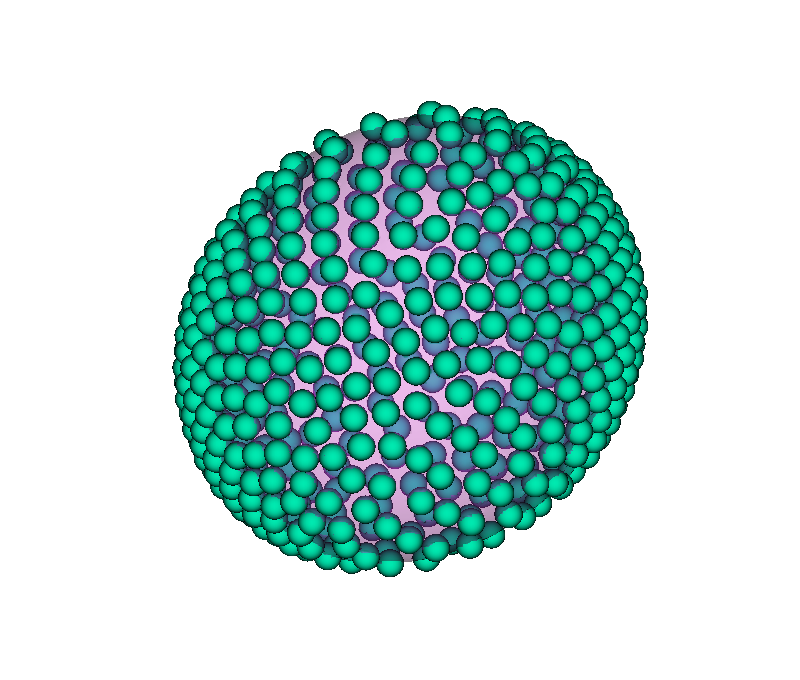} &
\includegraphics[width=0.18\linewidth]{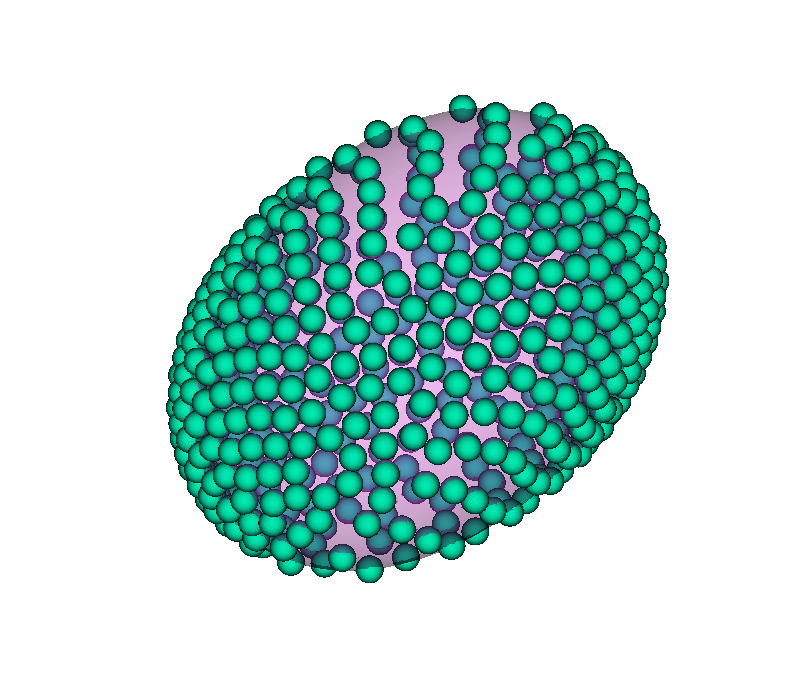} &
\includegraphics[width=0.18\linewidth]{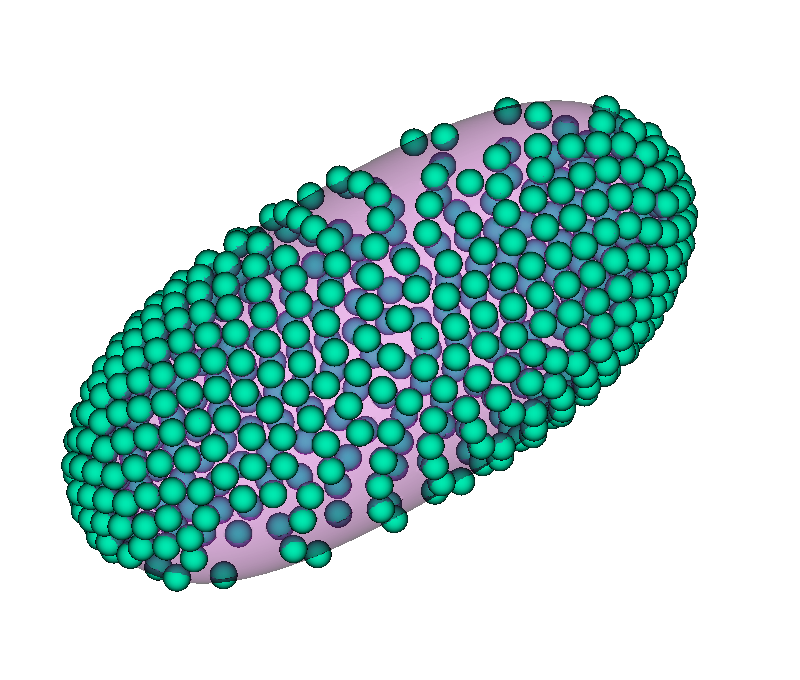} \\
& \revisedtext{$\mathrm{Ca}^{\mathrm{eff}} = 0.04$} & {$\mathrm{Ca}^{\mathrm{eff}} = 0.08$} & {$\mathrm{Ca}^{\mathrm{eff}} = 0.12$} \\
\end{tabular}
\end{center}
\caption{Side-view examples of deformed droplets, for various particle coverage
fractions: a) $\chi = 0.00$, b) $\chi = 0.27$ and c) $\chi = 0.55$. Although increasing $\chi$ from $0$ to $0.27$ does not strongly
change the deformation of the droplet, the particles prefer to stay in the middle of the channel where
the shear flow is weakest. The particles also exhibit
tank-treading-like behaviour: they move around the interface
following the shear flow (reprinted from~\cite{bib:jens-frijters-2012a} - reproduced by permission of The Royal Society
of Chemistry).}
\label{fig:pictures-sheared-particles}
\end{figure}
The fluid-fluid
interaction strength is held fixed at $g_{br} = 0.10$. The particles
have a radius of $r_p = 5.0$ and are neutrally wetting ($\theta_p = 90^\circ$).
We choose their mass to be $m_p = 524$, i.e. identical to the fluid mass density. The
introduction of finite-sized particles introduces a lower bound on how small
the simulation volume can be to accomodate enough particles on the interface
and to avoid finite-size effects. For this reason, the simulation volume is
chosen to be $n_x = n_y = 256$, $n_z = 512$, with an initial
droplet radius of $R^{\mathrm{init}}_d = 0.3 \cdot n_x = 76.8$, still keeping
it as small as possible to avoid excessive calculation time. The number of
particles is varied as $n_p = 0$, $128$, $256$, $320$, $384$, $446$ and $512$,
which results in a surface coverage fraction of $\chi = 0$ up to $\chi = 0.55$.
The capillary number is changed by changing the shear rate. Some
examples of the deformations thus realised are shown in
\figref{pictures-sheared-particles}, for $\mathrm{Ca}^{\mathrm{eff}} = 0.04$,
$0.08$, $0.12$ and $\chi = 0.0$ (a), $\chi = 0.27$ (b) and $\chi = 0.55$ (c).

When a droplet is sheared, its interfacial area increases due to the resulting deformation,
and more space is available for the particles to move freely over the interface (cf.~\figref{pictures-sheared-particles}).
However, even for high particle coverages and shear rates, detaching
particles from the interface remains practically impossible. The particles are
swept over the interface with increasing velocity as they move away from the
center plane of the system and up the shear gradient. If the particles would
not be affected by the shear flow, they would prefer to occupy interface with
high local curvature as can be explained by a geometrical argument: the
interface removed by a spherical particle at a curved interface is larger than
the circular area removed from a flat interface, and this effect gets stronger
as curvature increases. This explains why in this dynamic equilibrium, most
particles can be found at the tips of the droplet. This can be observed in
\figref{pictures-sheared-particles}~b) at high capillary number.

Even though the overall structure of the particles on the droplet interface
remains stable over time, individual particles move over the interface,
performing a quasi-periodic motion. However, this effect is qualitatively
different from the tank-treading behaviour observed in, for example,
vesicles~\cite{bib:kaoui-harting-misbah:2011}. While tank-treading vesicles are
characterized by a constant frequency of rotation for all points on their
membrane, this is not the case for the particle covered droplets: The
rotational frequency is not constant for all particles, instead showing a
dependence on the position of the particle, normal to the shear plane. The deformation is highest at the
center of the droplet surface, giving particles greater options for mobility that are
also better-aligned with the shear flow, leading to increased particle
velocities. 

\subsection{Droplet deformation and inclination}
\label{ssec:deformation-inclination-results}
\begin{figure}
\centerline{
\includegraphics[height=0.32\linewidth]{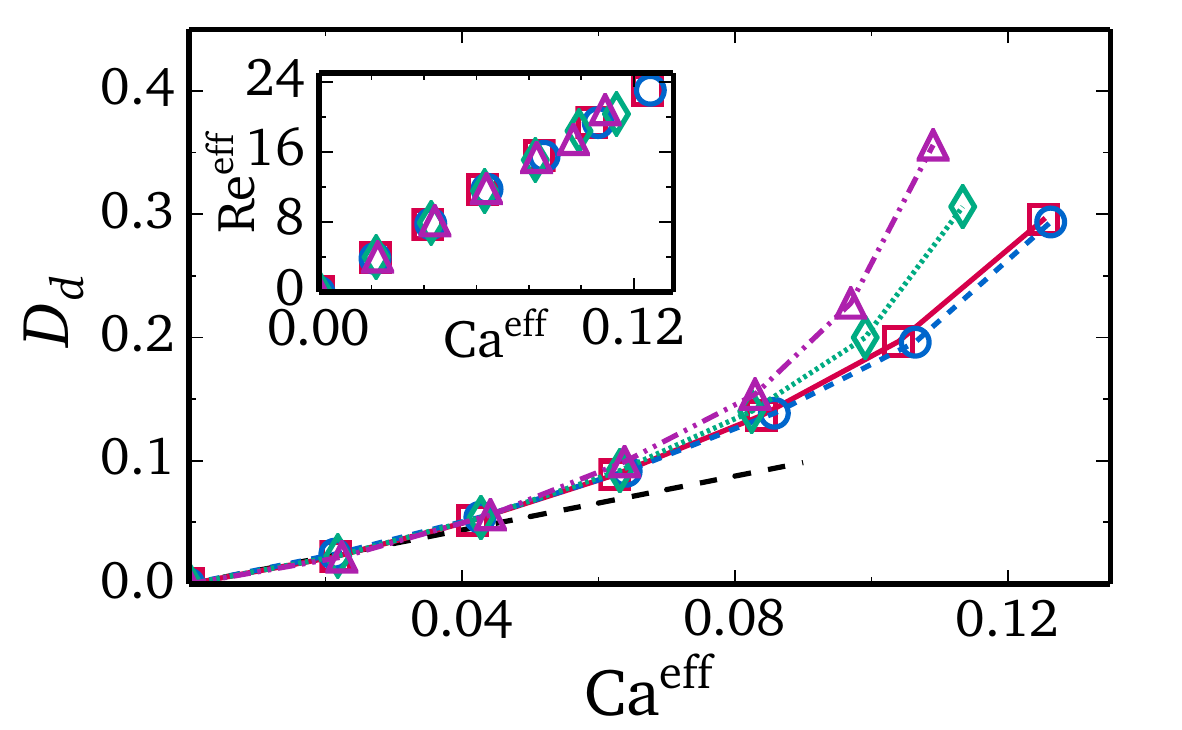}
%}
%\centerline{
\includegraphics[height=0.32\linewidth]{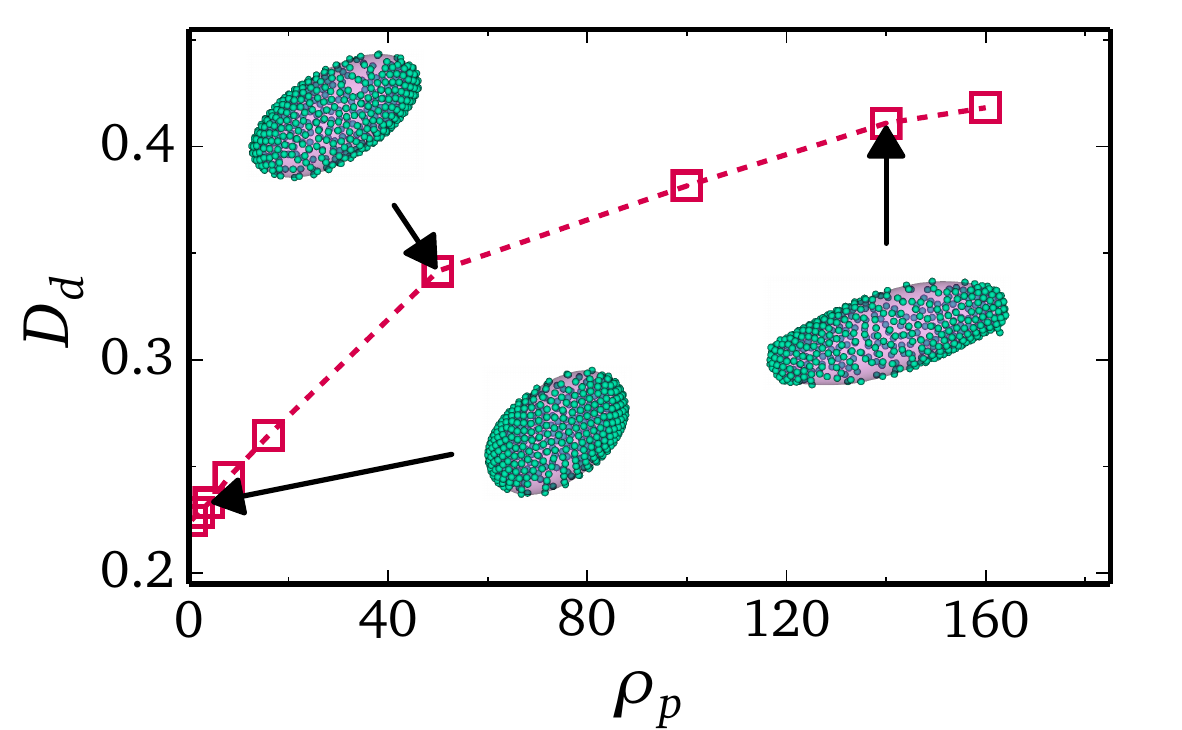}}
\caption{Left: The main plot shows the deformation parameter $D$ as a function of the effective
capillary number $\mathrm{Ca}^{\mathrm{eff}}$ for various degrees of droplet
interface particle coverage fraction $\chi$: $\chi = 0$ (red squares), $\chi = 0.27$ (blue circles), $\chi = 0.41$ (green diamonds), and $\chi = 0.55$ (purple triangles).
The effect of adsorped particles is very weak for low $\chi$, becomes noticeable at $\chi > 0.4$. Taylor's law is reproduced for small $\mathrm{Ca}$ (dashed
line). The inset depicts that the Reynolds number scales linearly with the capillary number.
Right: The deformation parameter $D$ is shown as a function of the rescaled
mass of the particles $m_p^* = m_p/m_p^0$, where $m_p^0 = 524$ is defined by
setting the density of the particle to 1. The particles have a radius $r_p =
5.0$, their coverage fraction is $\chi = 0.55$ and the capillary number is
$\mathrm{Ca}^{\mathrm{eff}} = 0.1$. Snapshots of the droplets are included,
showcasing the deformations of the droplet. The inertia of the heavier
particles causes additional deformation as they drag the droplet interface in
the direction of the shear flow 
(reprinted from~\cite{bib:jens-frijters-2012a} - reproduced by permission of The Royal Society
of Chemistry).}
\label{fig:plot-sheared-particles-deformation}
\end{figure}
%\label{fig:plot-deformation-mass}

The linear dependence of the deformation parameter $D$ on the effective
capillary number introduced above is recovered in our simulations for 
low capillary number and low particle
coverage
(cf.~\figref{plot-sheared-particles-deformation}, left). When the coverage fraction
grows beyond $\chi > 0.40$ the deformations in this regime increase with
increasing $\chi$ and constant capillary number. 

Combining Eqns.~\ref{eq:Ca-eff} gives a relation between the capillary and Reynolds number:
\begin{equation}
  \label{eq:Re-Ca}
  \mathrm{Re}^{\mathrm{eff}} = \sigma \left( \frac{\rho_m R_d}{\mu_m^2} \right) \mathrm{Ca}^{\mathrm{eff}}
  \hbox{.}
\end{equation}
As we change the capillary number explicitly by changing the shear rate, the
Reynolds number is proportional to the capillary number for a fixed value of
the surface tension. Inertial effects increase the deformation, thus the
deformations at high capillary number are higher than predicted by the linear
relation of Taylor.
Furthermore, since the nanoparticles do not affect the surface tension, all
curves of the Reynolds number versus capillary number have the same slope (cf. inset
of~\figref{plot-sheared-particles-deformation}(left) and \eqnref{Re-Ca}). This
implies that the increased deformation in the case of added nanoparticles is
not caused by changes in inertia of the fluids. On the other hand, the inertia
of the particles themselves plays a decisive role here. We have investigated
the dependence of the droplet deformation on the size and mass of the
particles. Particle radii have been varied between $4.0 \le r_p \le 10.0$ and
at $\mathrm{Ca}^{\mathrm{eff}} = 0.1$ this has led to only a small change in
$D$. Yet, changing the mass of the particles directly has a substantial effect.
We have varied the mass of the particles over two orders of magnitude, as shown
in \figref{plot-sheared-particles-deformation}(right). $\chi = 0.55$ and
$\mathrm{Ca}^{\mathrm{eff}} = 0.1$ are kept constant and we have rescaled the
mass scale with the reference mass: $m_p^* = m_p / 524$. The particles are
accelerated as long as they are on the part of the droplet interface that
experiences a shear flow at least partially parallel to the particle movement.
Eventually, particles have to ``round the corner'' and are forced to move
perpendicular to or even antiparallel to the shear flow. The increased inertia
of heavier particles makes it more difficult to change the movement of these
particles, leading to a situation where the droplet interface is in fact
initially dragged farther away in the direction of the shear flow instead. This
process is balanced by the surface tension as the surface area increases. This
then explains the increase of deformation with increasing particle mass. As our
deformation is increased substantially, the system size limits the deformation
we can induce. Therefore, the values presented here are underpredictions of the
actual effect of increased mass at high deformations, and might indeed hide a
breakup event.

\section{Meso-resolved Particle-Fluid Method}

This section is based on Refs.\ \cite{robinson2013ijmf,robinson2013sph,robinson2013dispersion},
where a meso-scale solver for particles dispersed in a fluid was
presented. Both phases are tackled by particle-based methods
\textemdash{}which facilitates their coupling\textemdash{}where
the fluid has a resolution somewhat larger than the discrete phase.

The SPH-DEM formulation is applied to several test cases
for verification and validation, before it is used to predict a dispersion experiment
with free-surfaces involved, which shows the quality of this particle-based method.
The agreement between model and simulation is very good, however, the feature of
lift-off of the packed bed is not reproduced yet. This can have various reasons,
since some features
like the interstitial gas, the surface tension and the polydispersity of the grain-packing,
are not implemented yet and are subject of ongoing research.

\subsection{The Smoothed Particle Hydrodynamics -- Discrete Element Method}
%The mesoscale simulation method presented in this section
%couples Smoothed Particle Hydrodynamics (SPH) and the Discrete
%Element Method (DEM) and enjoys the flexibility of meshless methods,
%such as being capable to handling free surface flows or flow around
%complex, moving and/or intermeshed geometries~\cite{robinson}.
%We use this method to simulate three different sedimentation test
%cases and compare the results to existing analytical solutions. The
%grain velocity in single particle sedimentation compares well (< 2\%
%error) with the analytical solution as long as the fluid resolution is
%coarser than two times the particle diameter. The multiple particle
%sedimentation problem and Rayleigh Taylor Instability (RTI) also
%perform well against the theory, but it is found that the method is
%susceptible to fluid velocity fluctuations in the presence of high
%porosity gradients. These fluctuations can be damped by the addition
%of a dissipation term, which has no effect on the terminal velocity
%but can lead to slower growth rates for the RTI.
%
We present a discrete particle method (DPM) based on the coupling of Smoothed
Particle Hydrodynamics (SPH) for the fluid phase and the discrete element method
(DEM) for the solid particles. This resuls in a purely particle-based simulation
method and enjoys the intrinsic flexibility as a primary advantage over
existing grid-based DPMs.
The model described in this section is well suited for applications
involving a free surface, including (but not limited to) debris flows,
avalanches, landslides, sediment transport or erosion in rivers and beaches,
slurry transport in industrial processes (e.g.~SAG mills) and liquid-powder
dispersion and mixing in the food processing industry
\cite{robinson2013sph}.
%\subsection{SPH Fluid Phase}\label{sec:SPH}
Here we briefly describe the governing SPH equations for the fluid phase, based
on the locally averaged Navier-Stokes equations (LANSEs) \cite{anderson67fluid};
for more details see Ref.\ \cite{robinson2013ijmf}.

First,
we define a smooth porosity field by smoothing out the DEM particle's volumes
according to the SPH interpolation kernel $W_{aj}(h) = W(r_a-r_j,h)$
\begin{equation}\label{eq:epsilonCalculation}
\epsilon_a = 1 - \sum_{j} W_{aj}(h) V_j,
\end{equation}
where $V_j$ is the volume of DEM particle $j$. For readability, sums over SPH
particles use the subscript $b$, while sums over surrounding DEM particles use
the subscript $j$.
To calculate the continuity and momentum equations in the LANSEs, we first define a superficial fluid density $\rho$ equal to the intrinsic fluid density scaled by the local porosity
$\rho=\epsilon \rho_f$.
Substituting the superficial fluid density into the averaged continuity and
momentum equations reduces them to the normal Navier-Stokes equations.
Therefore, our approach is to use the weakly compressible SPH
equations with variable $h$ (resolution/smoothing length) terms
\citep{monaghan05SPH,robinson11direct,price12smoothed}
and adding fluid-particle drag terms
(as specified below). %SL changed this sentence seriously - pls. check
%
%SL in the following I removed empty lines to make it more dense and avoid
% indents etc.
%
The rate of change of the superficial density then becomes
\begin{align} \label{Eq:changeInDensity}
\frac{D\rho_a}{Dt} = \frac{1}{\Omega_a}\sum_b m_b \mathbf{u}_{ab} \cdot \nabla_a W_{ab}(h_a), \notag \\
\Omega_a = 1 - \frac{\partial h_a}{\partial \rho_a} \sum_b m_b \frac{\partial W_{ab}(h_a)}{\partial h_a},
\end{align}
%MR
where $\mathbf{u}_{ab} = \mathbf{u}_a-\mathbf{u}_b$ is the difference in velocity between SPH-particles $a$ and $b$ and $\Omega_a$ is a correction factor due to the variable $h$.
%SL
Second, the acceleration of SPH-particle $a$ is given by
\begin{align}
\label{Eq:sphJustPressureForce}
\frac{d\mathbf{u}_a}{dt} & =
      -\sum_b m_b \bigg[ \left ( \frac{P_a}{\Omega_a \rho_a^2} +
      \Pi_{ab} \right ) \nabla_a W_{ab}(h_a) + \notag\\
& ~~~~~~~~~~~~~~~~~~ \left ( \frac{P_b}{\Omega_b \rho_b^2} + \Pi_{ab} \right )
               \nabla_a W_{ab}(h_b) \bigg]  + \mathbf{f}_a/m_a \,,
\end{align}
%SL in the right column of page 2 you refer to the gradients in this equation,
%   but they are not evident here - maybe change the reference to  grad P and div tau
%   such that you refer more clearly to this equation?
% OR: add the gradient terms into Equation (3) explicitly - we have just enough space.
%
where $\mathbf{f}_a$ is the coupling force on the SPH particle $a$ due to the
DEM particles. The term $\Pi_{ab}$ models the effect of the viscous
stress tensor and is calculated here using the term proposed
in Ref.\ \cite{monaghan05SPH}
\begin{equation}\label{Eq:monaghansViscousTerm}
\Pi_{ab} = - \alpha \frac{u_{sig} u_n }{2 \overline{\rho}_{ab} |\mathbf{r}_{ab}|} \,,
\end{equation}
%where XXX.
%SL pls. explain the symbols here in 1-2 lines max.
%SLnew
where $\overline{\rho}_{ab}$ and $\mathbf{r}_{ab}$ are the average density and
the distance between the centers of two SPH particles $a$ and $b$,
$u_n$ is the normal component of the relative velocity, 
and $u_{sig}= c_s + u_n/|\mathbf{r}_{ab}|$ is a signal velocity at which information propagates between the particles. The (numerical) sound speed is given by $c_s$ and $\alpha$ is a numerical pre-factor.

Third,
the fluid pressure in Eq. (\ref{Eq:sphJustPressureForce}) is calculated using
the weakly compressible equation of state where the reference density $\rho_0$ is scaled
by the local porosity to ensure that the pressure is slowly varying with porosity as
%SL changed text - pls. check
%
\begin{equation}\label{Eq:sphEquationOfState}
P_a = B \left ( \left ( \frac{\rho_a}{\epsilon_a \rho_0} \right )^\gamma - 1 \right ),
\end{equation}
%MR
where $B=100 \rho_0 u^2_m / \gamma$ is a scaling factor that is free a priori. It is set using a maximum velocity $u_m$, in order to ensure that density fluctuations (due to the weakly compressible SPH formulation) are below 1\%.
%
%JENS: WHAT IS "B"?
The smoothing length $h_a$ varies according to the superficial density (and hence with the porosity) and is calculated by $h_a = 1.5 ( {m_a}/{\rho_a} )^{1/3}$.
%
%\begin{equation}\label{Eq:variableh}
%h_a = \sigma \left ( \frac{m_a}{\rho_a} \right )^{1/3},
%\end{equation}
%
%where $\sigma = 1.5$.

%\subsection{DEM Solid Phase}\label{sec:DEM}
Finally,
the suspended particles are introduced as follows:
Given a DEM particle $i$ with position $\mathbf{r}_i$, the equation of motion
is
\begin{equation}
   m_i \frac{d^2 \mathbf{r}_i}{dt^2} = \sum_j \mathbf{c}_{ij} + \mathbf{f}_i +  m_i\mathbf{g},
\end{equation}
where $m_i$ is the mass of particle $i$, $\mathbf{c}_{ij}$ is the contact force
between particles $i$ and $j$ (acting from $j$ to $i$) and $\mathbf{f}_i$ is the fluid-particle coupling
force on particle $i$. For the simulations presented below, we have used the linear spring dashpot
contact model
\begin{equation}
\mathbf{c}_{ij} = -(k \delta -\beta \dot{\delta})\mathbf{n}_{ij},
\end{equation}
where $\delta$ is the overlap between the two particles and $\mathbf{n}_{ij}$ is the
unit normal vector pointing from $j$ to $i$.
The force on each solid particle by the fluid is \citep{anderson67fluid}
\begin{equation}\label{Eq:demCouplingForce}
\mathbf{f}_i = V_i (-\nabla P + \nabla \cdot \mathbf{\tau})_i + \mathbf{f}_d(\epsilon_i,\mathbf{u}_s),
\end{equation}
where $V_i$ is the volume of particle $i$. The first two terms model the
effect of the resolved fluid forces (buoyancy and shear-stress) on the
particle. The fluid pressure gradient and the divergence of the stress tensor are already calculated
%SL this ref to Eq. 3 is not clear ... pls. improve a little ... thanks
in Eq. (\ref{Eq:sphJustPressureForce}) and are evaluated at each solid particle,
using a Shepard corrected SPH interpolation~\citep{robinson2013ijmf}.

The force $\mathbf{f}_d$ models the drag effects of the unresolved
%SL fluctuations are not clear here
fluctuations in the fluid variables and is calculated from the local
porosity $\epsilon_i$ and the superficial velocity $\mathbf{u}_s = \epsilon_i (\mathbf{u}_f-\mathbf{u}_i)$. These two values are calculated at each DEM particle position,
again using a Shepard corrected SPH interpolation. For the results in this paper we use both the simple Stokes drag force and a more general drag law proposed by Di Felice \citep{difelice94voidage}.

The coupling force on SPH particle $a$ is determined by a weighted average of
the fluid-particle coupling force on the surrounding DEM particles.
\begin{equation}\label{Eq:SPHCoupleForce}
\mathbf{f}_a = - \frac{m_a}{\rho_a} \sum_j \frac{1}{S_j} \mathbf{f}_j W_{aj}(h_c),
\end{equation}
where $\mathbf{f}_j$ is the coupling force calculated for each DEM particle
using Eq. (\ref{Eq:demCouplingForce}) and $S_j = \sum_b{\frac{m_b}{\rho_b} W_{jb}(h_c)}$  is a correction factor to guarantee equal and opposite forces between the two phases.

\subsection{Verification case 1: Single Particle Sedimentation}\label{sec:SPS}

The first test case models a single particle sedimenting (SPS) in a 3D fluid
column under gravity \cite{robinson2013dispersion}. The water column has a height of $H=0.006 \mathrm{m}$, and
the bottom has a no-slip boundary, while the boundaries in the $x$ and $y$ directions
are periodic with a width of $w=0.004\text{ m}$.
Gravity acts in the negative $z$ direction, so that the single DEM particle, initialised
at $z=0.8H$, with diameter $d = 10^{-4} \text{ m}$ and density $\rho_p = 2500\text{ kg/m}^3$
starts to fall at $t=0$\,s. Note that before the SPS starts,
for the initial conditions of the simulation, the position of the DEM particle is
fixed and the SPH fluid particles are allowed to reach their hydrostatic equilibrium
positions, which are not necessarily the same as their input positions.

\begin{figure}
\centering
\includegraphics[height=0.47\textwidth, angle=-90]{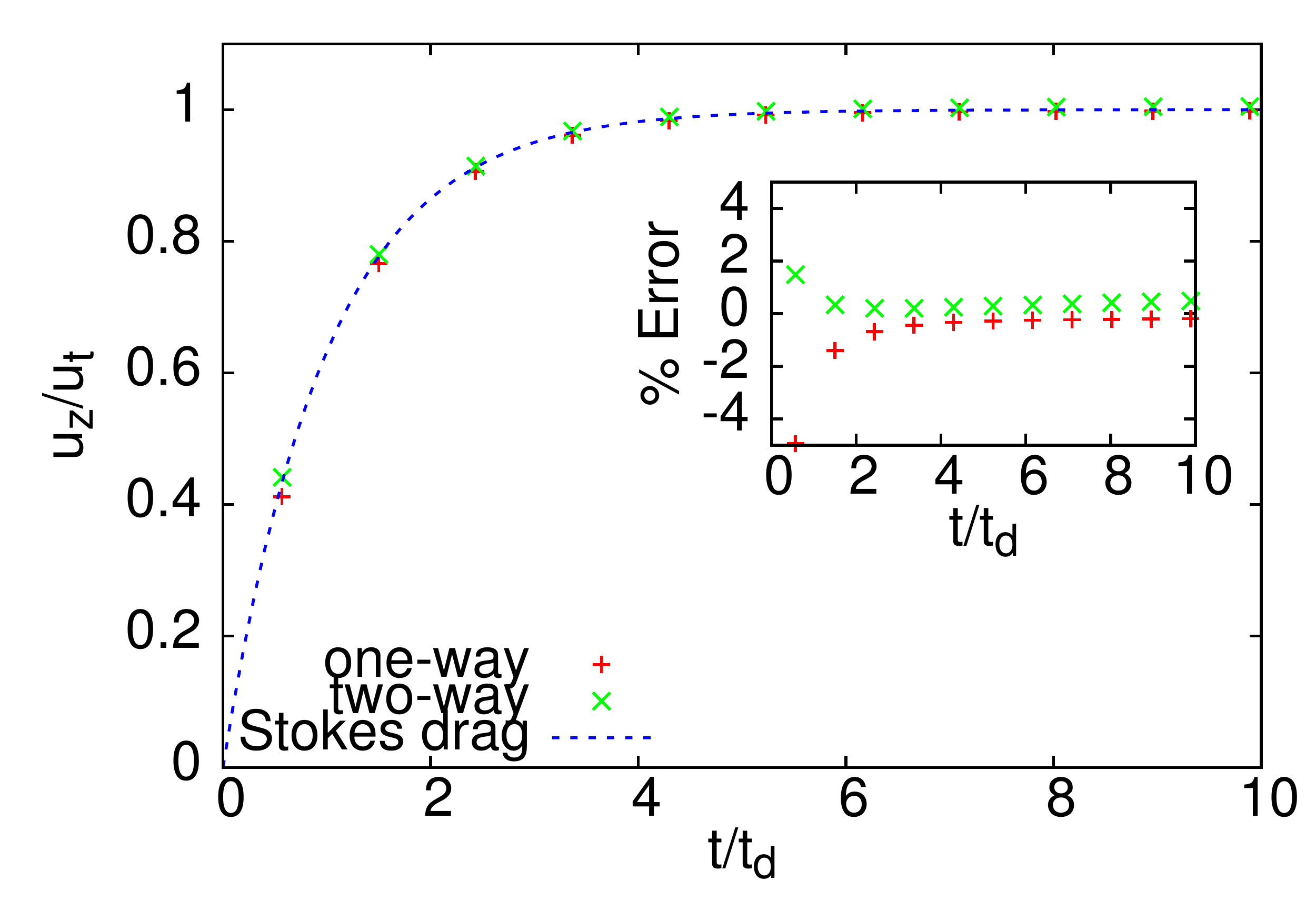}
\includegraphics[height=0.45\textwidth,angle=-90]{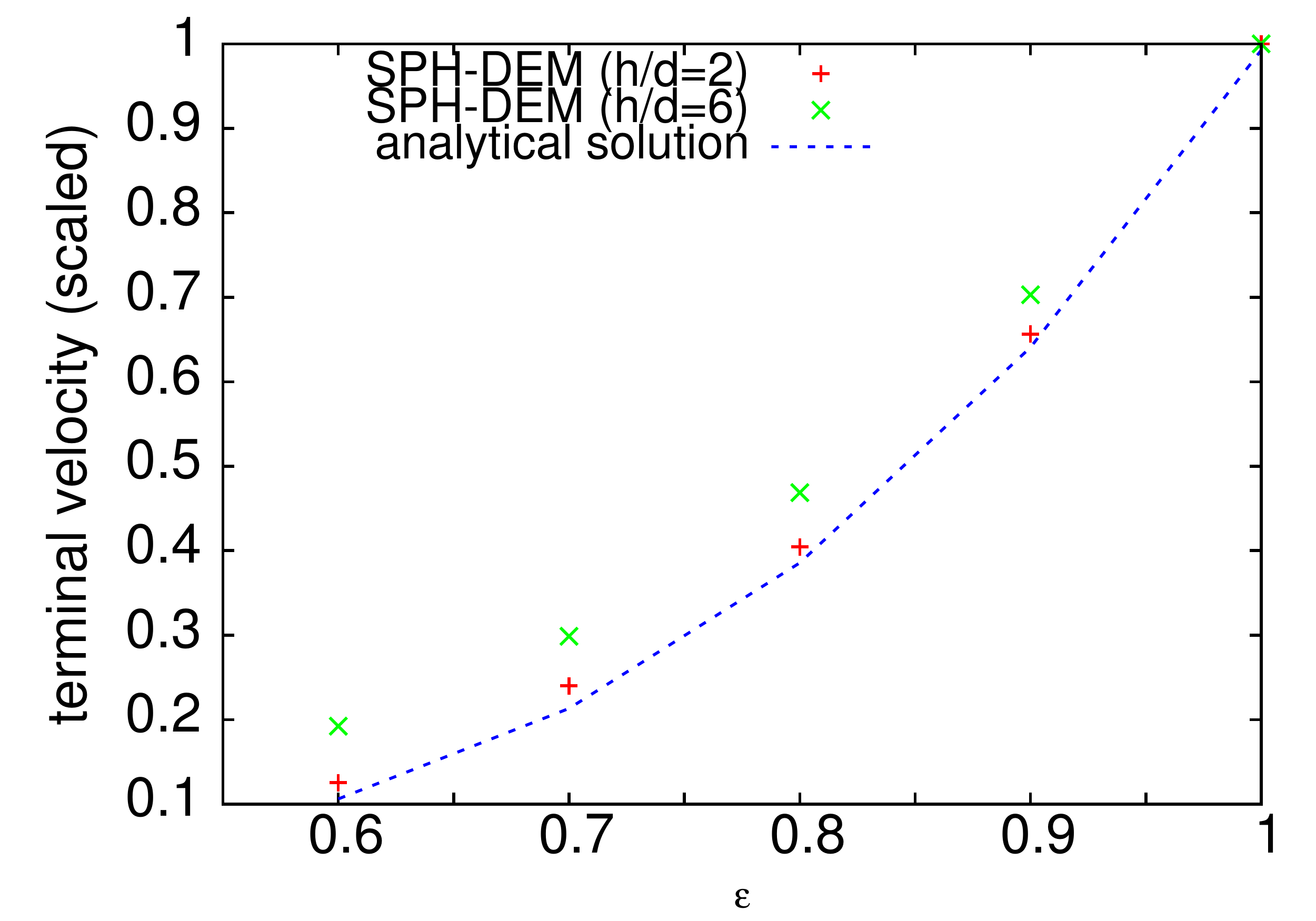}
\caption{Left: Sedimentation velocity for a single particle in water falling
from rest with both one-way and two-way coupling. The dashed line is the
theoretical result integrating Stokes law.  The $y$-axis shows the particle
vertical velocity scaled by the expected terminal velocity $u_t=|\mathbf{u}_t|$
and the $x$-axis shows time scaled by the drag relaxation time $t_d$. The inset
shows the percentage error between the SPH-DEM and the theoretically expected
trajectory.
Right:~Average terminal velocity (scaled by $|\mathbf{u}_t|$, the expected
terminal velocity of a single DEM particle) of the Constant Porosity Block
(CPB) in water for varying porosity and $h/d = 2$ and $6$, compared to the
analytical solution (dashed line).}
\label{fig:SPSoneway_water}
\end{figure}

In Figure \ref{fig:SPSoneway_water}(left) the evolution of a DEM particle's vertical
speed in water is shown for one-way and two-way coupling and for a reference
fluid with parameters corresponding to water. %We have performed similar
%simulations (data not shown) with fluids corresponding to air and a
%water-glycerol mixture~\cite{robinson2013dispersion}.
The SPH-DEM results
reproduce the analytical velocity curve within 0.3-1\% error besides
short-lived higher deviations at the initial onset of motion (approx 5\%).
One of the key assumptions of the SPH-DEM method (and any fluid-particle method
that uses an unresolved fluid phase) is that the fluid resolution is
sufficiently greater than the DEM particle diameter $d$. This ensures a smooth
porosity field calculated via Eq.\ (\ref{eq:epsilonCalculation}). By varying the
fluid resolution $h$, we have found that accurate results are achieved as long
as $h \geq 2d$.

\subsection{Validation case 2: Sedimentation of a Constant Porosity Block}\label{sec:CPB}
The second test case (CPB) follows the sedimentation of a rigid porous block
made up of a regular grid of DEM particles that cannot move relative to each
other \cite{robinson2013dispersion}. The porosity $\epsilon$ of the block determines the particle separation.
As the block falls, the fluid is displaced and flows upward through the block,
affecting the terminal velocity. All these simulations use the Di Felice drag
law, which is one possibility to incorporate the effects of neighbouring particles
on the drag force.

Varying the porosity of the CPB allows us to evaluate the accuracy of the
SPH-DEM model at different porosities. Fig.~\ref{fig:SPSoneway_water}(right) shows
the scaled terminal velocity of the CPB over a range of porosities from
$\epsilon=0.6$ to $1.0$.  The results for two different fluid resolutions are
shown, $h/d=2$ and $6$.  The lower resolution leads to a systematically
increased terminal velocity due to a reduced drag at the edges of the block,
caused by an excessive smoothing of the porosity discontinuity by the large
width of the smoothing kernel, a feature not restricted to SPH-DEM but common
to any fluid-particle method that uses an unresolved fluid phase.  However,
reducing the fluid resolution to $h/d=2$ gives results better than 5\%
deviation (for the smallest $\varepsilon$) over the range of porosities tested.
More details and other fluids, as well as other resolutions are discussed in
Ref.\ \cite{robinson2013ijmf}, where also another test-case, the Rayleigh
Taylor Instability (RTI) is introduced and used to study moving particles in
the inverse settling set-up and the growth of the associated instability.
Instead of continuing this, in the next subsection, we show one example with
free boundaries, where the SPH-DEM method is at its best.

\subsection{Dispersion example: Effect of Inlet Flow Rate}

Ref.\ \cite{robinson2013sph} compares experiments and SPH-DEM simulations of the dispersion of poppy seeds contained in a cylindrical cell, by liquid injected from a $1$\,mm diameter inlet at the bottom of the cell. The cell diameter is $d_{\rm cell} = 23$ mm and the cell height $h_{\rm cell} = 30$ mm. The outlet occupies all the top surface of the cell and is permeable to the liquid, but not to the grains. The grain density is $\rho_p = 1160\text{ kg/m}^3$ and their average diameter is $d = 1.1$ mm. The total mass of grains in the cell is $4.0$\,g, which corresponds to a bed height of about $14$\,mm.

The cylindrical cell and the cylindrical inlet/outlet are modelled as no-slip boundaries in the SPH-DEM simulation, using a single layer of SPH repulsive force particles~\cite{monaghan94simulating}. The presence of the top filter is captured by including a top horizontal wall permeable to the fluid, but not to the grains. The DEM grains are allowed to settle into a packed arrangement before the inlet jet is turned on at $t=0$. At this point SPH particles are added to the lower inlet with a constant velocity and rate that matches the experimental inlet flow rate $Q_i$. The air is assumed to have no influence on the results and is modelled implicitly by the absence of SPH fluid particles.

\begin{figure}
\centering
\includegraphics[height=0.34\textwidth,angle=-90]{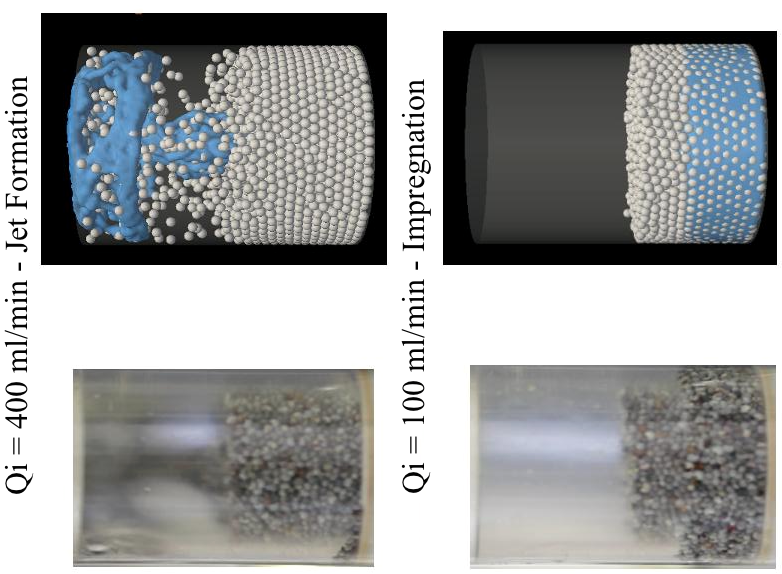}
\caption{
Dispersion of a dry granular bed of poppy seeds by a $1$\,mm central injection hole.
Left: Experimental dispersion patterns shortly after liquid injection starts.
Right: Pseudo-Three-Phase SPH-DEM simulations.
Grains are represented using white spheres and the liquid free-surface is colored in blue.
Top row: A $400$\,ml/min inlet flow rate generates a jet.
Bottom row: A slower $100$\,ml/min flow rate induces a bottom-up
impregnation~\cite{robinson2013sph}.
}
\label{fig:dispersion}
\end{figure}

In Ref.\ \cite{robinson2013sph}, a range of inlet flow rates $50 \le Q_i \le 600$\,ml/min was simulated. The behavior changes with increasing in-flow intensity.
%For strong flow $Q_i > 100$\,ml/min, the water jet fluidized a central column
%of the grain-bed and breaks through the top of the bed. Then, in a second phase,
%the cell filled with fluid penetrates the bed from the top to the bottom, and then
%fills the rest of the vessel upwards to the outlet.
and the behavior predicted by simulations closely matches that of the experimental
results, including the cut-off point at $Q_i = 100$\,ml/min.
Below this cut-off point the jet failed to fluidize the bed and the dispersion cell filled
with fluid from the bottom up to the outlet height. The movement of DEM grains was minimal
until the bed had been completely immersed, which qualitatively matches the
experimental results. 
Figure \ref{fig:dispersion} shows snapshots from the simulations.
Grains are represented using white spheres and the water free-surface is colored in blue.
For $Q_i = 100$\,ml/min the grains do not move significantly and the fluid free-surface
height linearly grows over time from the bottom to the top of the cell.
The fluid free-surface is approximately constant over the horizontal width of the cell.
For $Q_i = 400$\,ml/min the jet quickly breaks through the center of the grain bed,
dispersing a large number of grains throughout the cell. The jet is so strong that it
impacts on the top of the cell, whereas for smaller $Q_i$ the jet did not reach the top.
Once the jet breaks through, the cell fills with fluid from top to bottom until the
DEM grains are fully immersed.

\section{Conclusion}
We have summarized the advantages and disadvantages of a number of simulation
algorithms for particle-laden flows. Two recent state of the art
implementations based on a DEM algorithm for the particles coupled to a
multicomponent LBM or an SPH implementation for the involved solvents were
presented and and some specific applications were highlighted. To conclude,
there is no perfect candidate on the market, but the method of choice has to be
carefully selected depending on the physical problem. Of particular importance
are the relevance of long-range hydrodynamic interactions, particle
concentration and shape, the ratio of Brownian and advective forces, the role
of inertia, or non-hydrodynamic particle-particle interactions such as
electrostatic or van der Waals forces.   

%\bibliographystyle{abbrv-unsrt}
%\bibliography{jens-pub,bibliography,frijters,A7_references,MRobinson,new}

%\section{Acknowledgements}

\end{document}